\documentstyle[12pt,aaspp4,psfig]{article}

\newcommand{\anue}{\bar\nu_e}
\newcommand{\eps}{\varepsilon}
\newcommand{\Msun}{M_\odot}

\newcommand{\ltilde}
{~ \raisebox{-1ex}{$\stackrel{\textstyle <}{\sim}$} ~}

\begin{document}

\title{Future Detection of Supernova Neutrino Burst and Explosion
Mechanism}
\author{T. Totani}
\affil{Department of Physics, The University of Tokyo, Tokyo 113 Japan 
  \\ E-mail: totani@utaphp2.phys.s.u-tokyo.ac.jp}
\author{K. Sato}
\affil{Research Center for the Early Universe, The University of
  Tokyo, Tokyo 113 Japan}
\and
\author{H.E. Dalhed and J.R. Wilson}
\affil{Lawrence Livermore National Laboratory \\
7000 East Avenue, L-015, Livermore, CA94550, U.S.A.}

\begin{abstract}
Future detection of a supernova neutrino burst by large underground
detectors would give important information for the explosion mechanism
of collapse-driven supernovae. We studied the statistical analysis 
for the future detection of a nearby supernova 
by using a numerical supernova model and realistic Monte-Carlo 
simulations of detection by the Super-Kamiokande detector. 
We mainly discuss the detectability
of the signatures of the delayed explosion mechanism in the time evolution
of the $\anue$ luminosity and spectrum. For a supernova at 10 kpc
away from the Earth, we find that not only the signature is clearly
discernible, but also the deviation of
energy spectrum from the Fermi-Dirac (FD) distribution can be observed.
The deviation
from the FD distribution would, if observed, provide a test
for the standard picture of neutrino emission from collapse-driven supernovae.
For the $D$ = 50 kpc case, the signature of the delayed explosion
is still observable, but statistical fluctuation is too large to
detect the deviation from the FD distribution. 
We also propose a method for statistical reconstruction of the time 
evolution of $\anue$ luminosity and spectrum from data, by which
we can get a smoother time evolution
and smaller statistical errors than a simple, time-binning analysis.
This method is useful especially when the available number of events 
is relatively small, e.g., a supernova in the LMC or SMC.
Neutronization burst of $\nu_e$'s produces about 5 scattering events
when $D$ = 10 kpc and this signal is difficult to distinguish
from $\anue p$ events.

\end{abstract}

\newpage
\section{Introduction} 
Although the Type II (and Ib, Ic) supernovae are generally believed to be
associated with 
gravitational collapses of massive star cores at the end of their life,
the explosion mechanism of the collapse-driven supernova 
has not yet been clarified. 
Two possible scenarios are so far discussed:
the prompt (Wilson 1971; Hillebrandt 1982; Arnett 1983; Hillebrandt, Nomoto, \&
Wolff 1984; Baron, Cooperstein, \& Kahana 1985)
or delayed (Wilson 1985; Bethe \& Wilson 1985) explosion. 
If the prompt explosion obtains, the envelope of progenitor stars
is directly expelled on the time scale of $\sim$ 10 msec
by the shock wave, which is generated by the core bounce when
collapse has compressed the inner core
of massive stars up to supranuclear densities.
It is known that, however, this mechanism can work only
when the progenitor star is relatively small ($\sim 10 M_\odot$)
and equation of state is soft enough, otherwise the shock wave
loses its energy by photodissociation of heavy nuclei and neutrino
emission behind the shock front, and finally stalls.
Therefore the delayed explosion scenario,
in which the stalled shock is revived by the heating of neutrinos
from the nascent neutron star on the 
time scale of $\sim$ 1 sec, is considered to be likely.

Since electromagnetic waves cannot convey to us any information of 
dense and deep region relevant to the explosion mechanism, 
the detection of a neutrino
burst by large underground detectors is almost the only chance 
to get some observational clues for the explosion mechanism.
Supernova neutrinos have already been detected from the supernova
SN1987A, which appeared in the Large Magellanic Cloud (LMC),
by the two water \v{C}erenkov detectors, Kamiokande II 
(Hirata et al. 1987, 1988)
and IMB (Bionta et al 1987; Bratton et al. 1988). Although these detections
were epoch-making events, the small numbers of captured neutrinos
(11 for Kamiokande and 8 for IMB) were unfortunately too small to tell us
something about the explosion mechanism. 
However, an international network of second-generation neutrino
detectors is now emerging on the Earth, and a future event in our
Galaxy or Magellanice Clouds would give us much more detailed data
necessary to understand the explosion mechanism (see, e.g., Burrows,
Klein, \& Gandhi 1992).
Among such underground detectors, the Super-Kamiokande (SK) detector, which
is about 15 times larger than the Kamiokande, has started its
observation and would detect 5000--10000 $\anue$'s if a supernova
appeared in the Galactic Center (10 kpc away from the Earth)
(Totsuka 1992; Nakamura et al. 1994). Because the signatures of
the explosion mechanism
are seen in the time evolution of the neutrino luminosity
and spectrum, statistical reconstruction of the time
evolution, that requires a large number of events, is crucially important. 
From this viewpoint, the SK, to which we confine ourselves in this
paper, is the most 
suitable detector to probe the explosion mechanism because of
its largest detector mass and good energy resolution,
among the existing or planned detectors.

In this paper, we investigate the signatures of 
the delayed explosion mechanism which are observable in the time
evolution of the
$\anue$ luminosity and spectrum at the SK,
by using a numerical model of supernova neutrino emission and 
realistic Monte-Carlo simulations of the detection by the SK.
The simplest analysis is to set many bins
in the time coordinate and estimate the luminosity and average energy
in each bin. However, this analysis is not sufficient to reproduce
a smooth evolutionary curve and some information of detection time
is also lost. We propose an analysis method based on the maximum likelihood
method and cubic-interpolation, which gives natural and smooth
evolution without loss of time information, and test this method by
using MC data generated from the numerical supernova model.

Because the supernova neutrinos are emitted thermally,
the Fermi-Dirac (FD) distribution was generally assumed in previous
analyses for the SN1987A data (Loredo \& Lamb 1989, and references therein). 
However, it is known that because the neutrino opacity
changes with neutrino energy, the spectrum of neutrinos
deviates from the pure FD distribution with zero chemical potential.
The emergent neutrino spectrum can be considered to be a blackbody
radiation from a surface whose radius varies with neutrino energy,
and deficit in both the low and high energy range compared to
the pure FD distribution is seen (Bruenn 1987; Mayle, Wilson, \& Schramm
1987; Janka \& Hillebrandt
1989; Giovanoni, Ellison, \& Bruenn 1989; Myra \& Burrows 1990).
The rich statistics of the SK may allow us to discern this deviation,
and if observed, it would provide a verification for the
current theoretical picture of supernova explosion and neutrino emission.
Hence we also investigate whether we can see this deviation of 
neutrino spectrum from the FD distribution in future observations.

In \S \ref{section:simulation}, we describe the numerical 
model of a supernova explosion
used in this paper, and the features of the delayed explosion are 
discussed. The properties of the SK detector are summarized
in \S \ref{section:SK-MC}, and Monte-Calro data generation from the numerical
supernova model is also described. Statistical analysis for the
MC data and its results are presented in \S \ref{section:analysis}, 
and after some discussions in \S \ref{section:discussion}, we
summarize our results in \S \ref{section:summary}.

\section{Numerical Model of Supernova Explosion}
\label{section:simulation}
\begin{figure}
  \begin{center}
    \leavevmode\psfig{figure=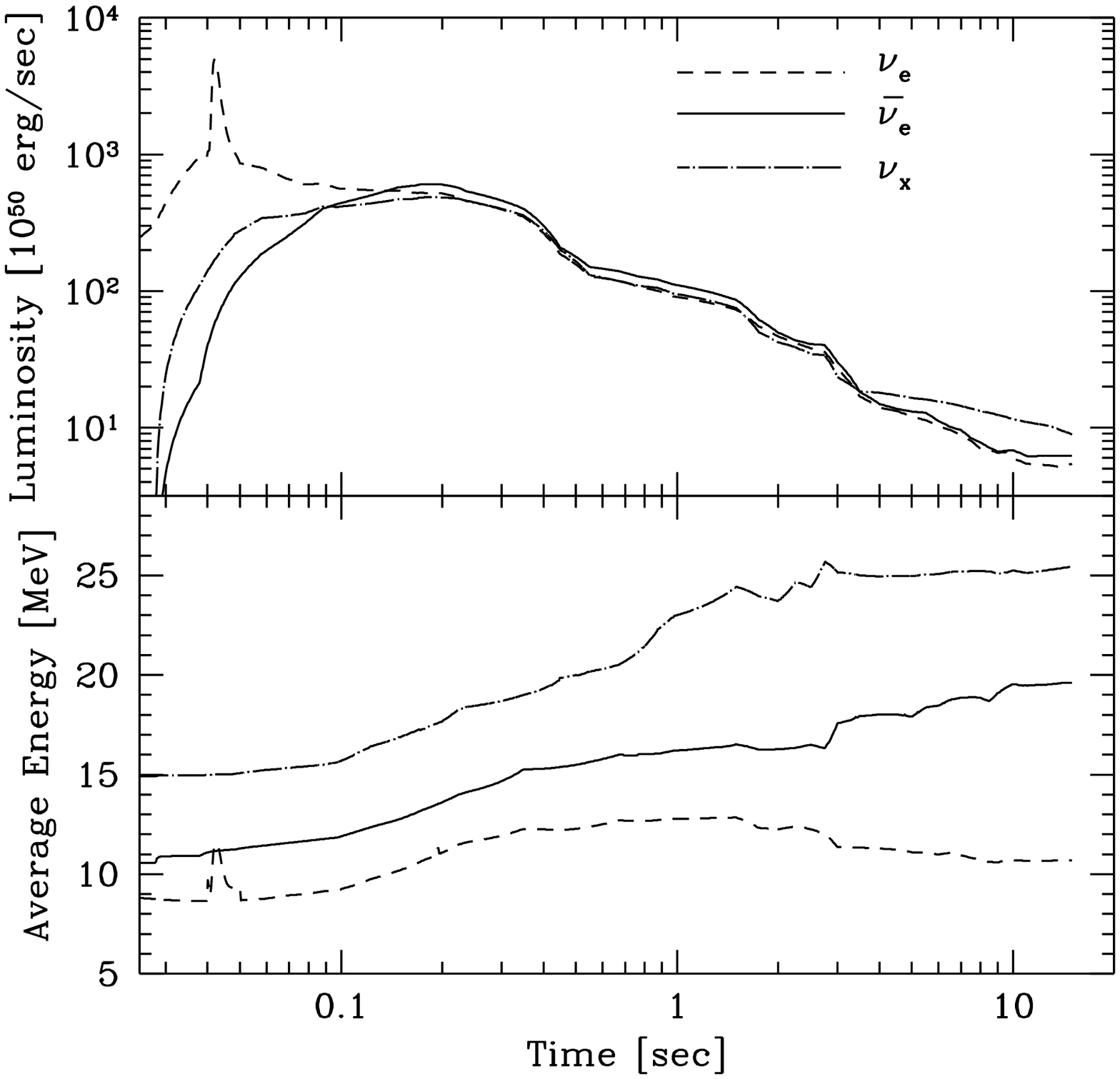,width=7cm}
  \end{center}
\caption{
Time evolution of neutrino luminosity and average energy
of the numerical supernova model used in this paper. The dashed
line is for $\nu_e$, solid line for $\anue$, and dot-dashed line
for $\nu_x$ (= each of $\nu_\mu, \nu_\tau, \bar\nu_\mu$, and
$\bar\nu_\tau$). The core bounce time is 3--4 msec before the
neutronization burst of $\nu_e$'s.}
\label{fig:WM-log}
\end{figure}

\begin{figure}
  \begin{center}
    \leavevmode\psfig%
    {figure=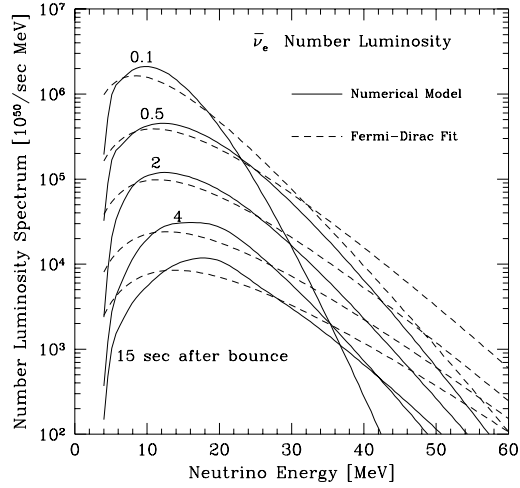,width=7cm}
  \end{center}
\caption{
Energy spectrum of $\anue$'s of the numerical supernova model
used in this paper. The time (after the bounce) is indicated
in the figure. The dashed lines are the Fermi-Dirac fits which have
the same luminosity and average energy with the numerical model.
The chemical potential is set to zero for the FD distribution.}
\label{fig:WM-spec}
\end{figure}

We use a result of neutrino emission based on a numerical simulation
of supernova explosion.
The simulation is performed with the numerical codes 
developed by Wilson and Mayle (Mayle 1985; Wilson et al 1986; 
Mayle, Wilson, \& Schramm 1987). 
The simulation is a model of the SN1987A, whose progenitor is a
main-sequence star of about 20 $\Msun$. 
The stellar configuration from which the explosion calculation
started was supplied by Woosley and Weaver (1991). This one dimensional
simulation is performed from the onset of the collapse to 
18 seconds after the core bounce in a consistent way, 
and the total energy emitted by this time is 2.9 $\times 10^{53}$ erg.
The emitted energy in $\anue$'s is 4.7 $\times 10^{52}$ erg and
the average energy of $\anue$'s is 15.3 MeV.
Figure \ref{fig:WM-log} shows the time evolution of
luminosity and average energy of this model, for 
$\nu_e, \anue$, and $\nu_x$. [The neutrino luminosity and spectrum
of supernova $\nu_\mu, \bar\nu_\mu, \nu_\tau$, and $\bar\nu_\tau$ 
(referred to $\nu_x$, hereafter) are almost the same.]
Figure \ref{fig:WM-spec} shows the snapshots of 
the spectral evolution of supernova $\anue$'s
in the form of differential number luminosity.
FD spectra which have the same luminosity
and average energy with zero chemical potential
are also shown by dashed lines in the figure,
and the deficit of both low- and high-energy neutrinos can be seen.
In the following analysis, FD distributions will be introduced
to provide a simple method of analyzing the possible observational
data and do not represent any expected spectral shape.
For generic features of supernova neutrino emission, see, e.g.,
Burrows et al. (1992). 

\begin{figure}
  \begin{center}
    \leavevmode\psfig%
    {figure=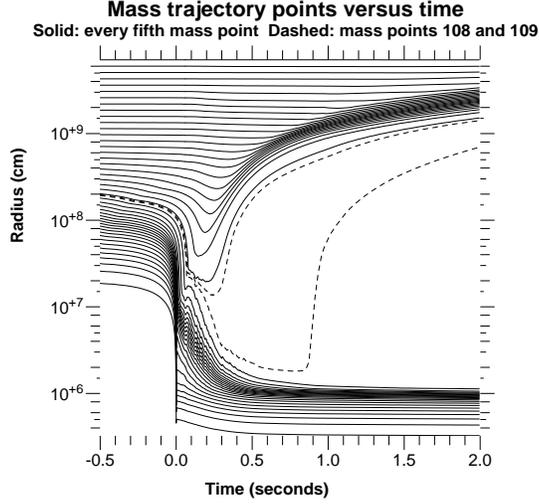,width=7cm}
  \end{center}
\caption{
Radius as a function of time for selected mass points of
the numerical supernova model used in this paper. Solid lines are
drawn for every fifth mass point, while the dashed lines are for
two succeeding mass points near the edge of the nascent neutron star
and ejected matter.}
\label{fig:mt}
\end{figure}

\begin{figure}
  \begin{center}
    \leavevmode\psfig{figure=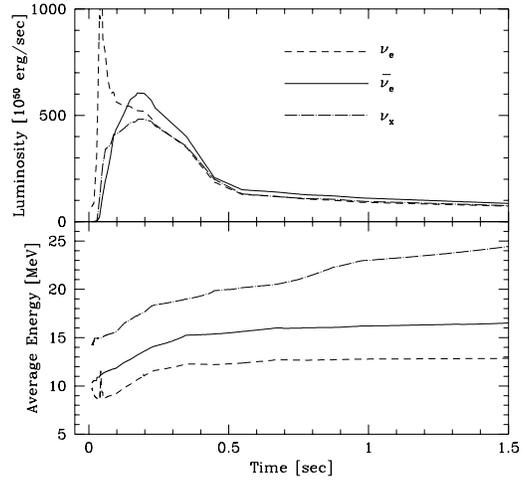,width=7cm}
  \end{center}
\caption{
The same as Fig. \protect\ref{fig:WM-log}, but for
the early phase in linear coordinate.}
\label{fig:WM-lin}
\end{figure}

Figure \ref{fig:mt} shows the radius of selected mass points as a
function of time for the present model.
This model explodes by the delayed explosion mechanism, and its
features are stamped onto the early evolution of neutrino luminosity and
spectrum as shown in Fig. \ref{fig:WM-lin}, in which the evolution of 
neutrino luminosity and average energy in the first $\sim 1$ second
after the core bounce are shown.
A characteristic of the delayed mechanism is a ``hump''
in the neutrino luminosity curve 
due to the accretion of matter onto the nascent
neutron star. [This ``hump'' is the same with the phenomenon previously
discussed in Burrows et al. (1992) as ``abrupt drop in luminosity curve''.]
The average energy of neutrinos stays low during the hump
because of the dense matter above the neutrino sphere which contributes
to the neutrino opacity. The average energy gradually increases during
this phase, corresponding to gradual decay of mass
accretion before the delayed explosion commences. Therefore this
gradual hardening of neutrino spectrum also gives another observable
signature of the delayed explosion. After the accretion
decays, the evolution time scale in both the luminosity and average
energy becomes much longer than that in the earlier phase.
The time duration of the hump in the delayed explosion 
is very hard to calculate
from first principles so the measurement of this duration is most
important for the understanding of the explosion process. 
On the other hand, if the prompt mechanism were viable, the envelope
would be expelled in $O$(10) msec after the core bounce and 
there would be no hump in the luminosity curve because of no matter
accretion. The neutrino spectrum would become harder just after the core
bounce and explosion, in contrast to the delayed explosion.
This mechanism is likely at work for relatively low mass stars.

In the present model of supernova explosion, 
this hump is clearly discernible corresponding to the matter accretion
during the first 0.5 [sec]. We concentrate on
these two features, i.e., the hump in neutrino luminosity and 
increase in average energy during the first $\sim$ 0.5 [sec] as
the signatures of the delayed explosion.
It should be noted, however, that these features of the delayed explosion
generally depend on the mass of initial iron cores or progenitor
stars. The more massive iron cores lead to the more complicated 
structure in time evolution of neutrino luminosity and spectrum.
In some calculations of the delayed explosion (Mayle 1985; Mayle,
Wilson, \& Schramm 1987), oscillatory behavior was observed
whose existence may depend on the initial structure of the star.

\section{Monte-Carlo Simulation of the SK detection}
\label{section:SK-MC}
The Super-Kamiokande detector (Totsuka 1992; Nakamura et al. 1994)
is a water \v{C}erenkov detector whose fiducial volume for a
supernova neutrino burst is 32,000 ton. We have performed Monte-Carlo
(MC) simulations for the detection of supernova neutrinos by this
detector, which give a data set of detection time and energy of events.
We use only the time and energy information of electrons or positrons,
and do not consider the directional information because positrons
are emitted almost isotropically in the dominant $\anue p$ reaction.
The following reactions are taken into account:
\begin{eqnarray} 
\anue + p & \rightarrow & e^+ + n \quad (\eps_\nu^{th} = 1.8 {\rm MeV})
  \quad ({\rm C.C.}), \label{eq:anue-p} \\
\nu_e \ (\anue) + e^- & \rightarrow & \nu_e \ (\anue) + e^- \quad 
  ({\rm C.C.+N.C.}), \\
\nu_x + e^- & \rightarrow & \nu_x + e^- \quad ({\rm N.C.}),  \\
\nu_e + {\rm ^{16}O} & \rightarrow & e^- + {\rm ^{16}F} 
  \quad (\eps_\nu^{th} = 15.4 {\rm MeV}) \quad ({\rm C.C.}),  \\
\anue + {\rm ^{16}O} & \rightarrow & e^+ + {\rm ^{16}N} 
  \quad (\eps_\nu^{th} = 11.4 {\rm MeV}) \quad ({\rm C.C.}), \label{eq:anue-O}
\end{eqnarray}
where C.C. and N.C. referred to charged and neutral current,
respectively, and $\eps_\nu^{th}$ is the threshold energy of neutrinos.
For the cross sections of $\anue p$ and $\nu_e (\anue) ^{16}$O
reactions, we referred to Vogel (1984) and Haxton (1987), respectively.
Although the cross sections of oxygen reactions are less certain
than others, this uncertainty hardly affects the following results
because of the much smaller number of events 
than that of the $\anue p$ reaction.
Recently Langanke, Vogel, \&
Kolbe (1996) pointed out that ($\nu, \nu' p \gamma$) and ($\nu, \nu' n
\gamma$) reactions on $^{16}$O constitute significant signals 
in the observed $\gamma$-ray energy range of $\ltilde$ 10 MeV, which are not
included in our MC code. The signal would
allow a unique identification of $\nu_x$, but simultaneously
be a serious noise against $\anue p$ events, in which we are interested
here, because the SK cannot distinguish $\gamma$-rays and positrons
in this energy range. In order to avoid this noise, we set the threshold energy
of positron energy, $\eps_e^{th} = 10$ MeV. Because the detection
efficiency of the SK is 100 \% above $\eps_e \sim$  8 MeV, 
where $\eps_e$ is energy of electrons or positrons, we do not
have to consider the detection efficiency. The energy resolution of the SK,
16 $(10 {\rm MeV} / \eps_e )^{1/2}$ (\%) (Nakamura et al. 1994), is taken
into account in our MC codes assuming the Gaussian distribution.
[For details of the MC data generation of water \v{C}erenkov
detectors, see, e.g., Krauss et al. (1992). Our MC simulation is
basically similar to that of Krauss et al.]

\begin{figure}
  \begin{center}
    \leavevmode\psfig{figure=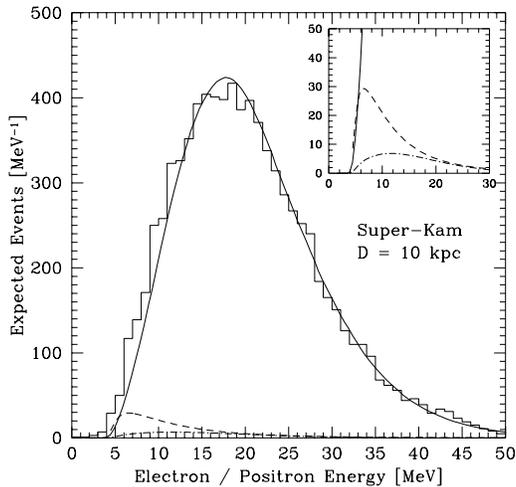,width=7cm}
  \end{center}
\caption{
Time-integrated energy distribution of election or positron 
events at the 
Super-Kamiokande detector for a supernova at 10 kpc away from the
Earth. The solid line shows the expected distribution of $\anue p$
events from the numerical supernova model. The dashed and dot-dashed
lines are for neutrino-electron scattering events 
(including all flavors of neutrinos) and $\nu_e (\anue) ^{16}$O
events, respectively. The histogram is an example of MC simulations
generated from the numerical supernova model, including all reaction
modes. The inset is a magnification of the lower energy region.}
\label{fig:dNde}
\end{figure}

\begin{figure}
  \begin{center}
    \leavevmode\psfig{figure=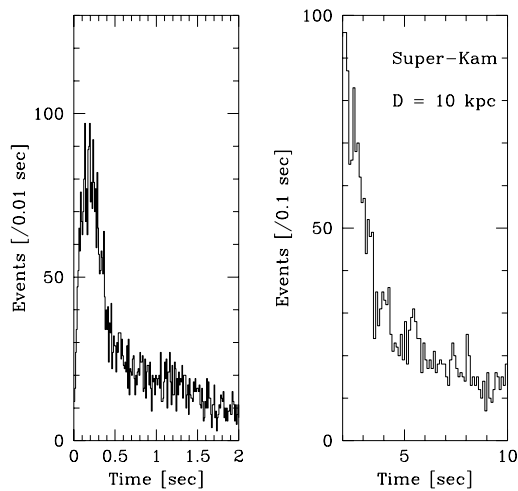,width=7cm}
  \end{center}
\caption{
Time histogram of an example of MC simulations of the
Super-Kamiokande detection generated from the numerical supernova
model. The distance to the supernova is set to 10 kpc. The time
of first event is set to zero.}
\label{fig:MC-time}
\end{figure}

The time-integrated, expected event distribution in positron or 
electron energy at the
SK is shown in Fig. \ref{fig:dNde} for each reaction mode, for the
case of a supernova at the Galactic center, i.e., $D$ = 10 kpc.
An example of the MC simulation is also shown as a histogram, which
includes all reaction modes. Because we
set $\eps_e^{th}$ = 10 MeV, the contamination of scattering events
and oxygen events to $\anue$p events is negligible. Figure \ref{fig:MC-time} 
shows the time histogram of events of this sample data set. 
The numerical supernova model
used here produces about 8300 events in the range of $0 \leq t \leq 18$ [sec], 
when $D$ = 10 kpc.
Most of these events are due to the $\anue p$ reaction.
The neutrino-electron scattering reactions produce about 200 events, and
the $\nu_e$ and $\anue$ absorptions into $^{16}$O produce about 100.
Since we cannot know the exact time of the onset of the collapse,
we set the time of the first event to zero. 

\begin{figure}
  \begin{center}
    \leavevmode\psfig{figure=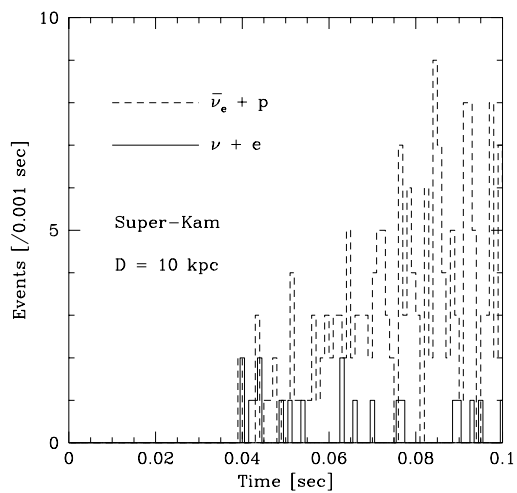,width=7cm}
  \end{center}
\caption{
The same as Fig. \protect\ref{fig:MC-time}, but for very early 
phase including the neutronization burst. The solid line for
the neutrino-electron scattering events of all neutrino flavors, 
and dashed line for events of $\anue p$ reactions.}
\label{fig:MC-time-early}
\end{figure}

The neutronization burst, namely, a strong peak in $\nu_e$ luminosity 
(reaching $\sim 5 \times 10^{53}$ erg/sec after 3--4 msec after the 
core bounce)
when the shock wave passes the neutrino sphere has attracted great attention 
of researchers
because its signal could be used for probing the electron neutrino mass or
probing the neutrino oscillation such as $\nu_e \leftrightarrow
\nu_\mu$ or $\nu_\tau$. It is also expected that this signal is
more energetic if the delayed explosion mechanism obtains rather than the
prompt mechanism (Burrows et al. 1992). However, the burst duration is  
$\sim$ a few msec and emitted energy as $\nu_e$'s is therefore
only several $\times 10^{51}$ erg. In addition, the cross section of $\nu_e
e$ scattering is about 1--2 orders of magnitude lower than
that of the $\anue p$ reaction, and hence it is doubtful that we can
clearly recognize the neutronization burst even in the case
of a Galactic supernova detected by the SK. During the neutronization
burst, the average energy of $\nu_e$'s is $\sim$ 11 MeV and 
expected number of 
scattering events is $\sim 2.7 (E_{\nu_e}/10^{51} {\rm erg})$.
Figure \ref{fig:MC-time-early} shows an example of the MC simulation for very
early events including the neutronization burst; the solid line
is for electron scattering events mainly due to $\nu_e$'s, while
the dashed line is for $\anue p$ events. Although the scattering
events are strongly forward peaked, it seems difficult to distinguish
the neutronization burst from $\anue p$ events when we take account
of the angular resolution of the SK ($\sim$ 30$^{\circ}$).

\section{Statistical Analysis for Reconstruction of $\anue$'
  Flux and Spectrum}
\label{section:analysis}
\subsection{Time-binning: the Simplest Analysis}
In the following part of this paper we consider reconstruction of 
$\anue$ flux and spectrum from $\anue p$ events. In order to
seek the signature of the delayed explosion mechanism, we have to
reconstruct the time evolution of $\anue$ luminosity and spectrum
during the first 1 sec after the bounce. The simplest procedure
to do this is to divide the time coordinate into many bins and 
estimate the luminosity and average energy in each time-bin.
Let $N_i$, $(\Delta t)_i$, and $(\bar\eps_e)_i$ be number
of events, width, and average energy of positrons or 
electrons in the $i$-th bin, in the energy range where analysis is performed.
As mentioned in \S \ref{section:SK-MC}, we take this energy range
as $\eps_e > \eps_e^{th}$= 10 MeV. In the following analysis,
we assume the Fermi-Dirac distribution with a single temperature
and zero chemical potential
for the simplicity. As discussed in Introduction, this assumption
is clearly oversimplification, but since we are interested in the
time evolution of $\anue$ luminosity and average energy, this
assumption does not affect the conclusion seriously. Note that
the analyzed MC data are generated by a numerical simulation of
multi-energy-group neutrino diffusion which does not assume any
shape of neutrino spectrum. This makes it possible for us to
investigate whether we can see the deviation of the neutrino spectrum
from the pure FD distribution by comparing the FD fits to energy
distributions of MC data.  We have also tried
a fit with time-evolving, nonzero chemical potential in FD
distribution, but it is found that we cannnot constrain strongly
the chemical potential,
if we set a rather high threshold, $\eps_e^{th}$= 10 MeV.

We can then easily estimate the luminosity $L_{\anue}^i$ and 
effective temperature $T_{\anue}^i$ in the $i$-th bin
by solving the following equations:
\begin{eqnarray}
\label{eq:bin1}
(\Delta t)_i \int_{\eps_e^{th}}^\infty d\eps_e \ 
\frac{d^2N (\eps_e; L_{\anue}^i, T_{\anue}^i)}{dtd\eps_e} &=& N_i \ , \\
\label{eq:bin2}
\int_{\eps_e^{th}}^\infty d\eps_e \ \eps_e 
\frac{d^2N(\eps_e; L_{\anue}^i, T_{\anue}^i)}{dtd\eps_e} &=& (\bar\eps_e)_i
\int_{\eps_e^{th}}^\infty d\eps_e \
\frac{d^2N(\eps_e; L_{\anue}^i, T_{\anue}^i)}{dtd\eps_e} \ ,
\end{eqnarray}
where $d^2N/dtd\eps_e$ is the expected event rate per unit time
per unit energy, which is generally given in the following form:
\begin{equation}
\frac{d^2N}{dtd\eps_e} = \sum_l \sum_m \int_0^\infty d\eps_\nu \
\frac{dF^l(\eps_\nu)}{d\eps_\nu} \ \frac{d\sigma^{lm}(\eps_\nu,
\eps_e)}{d\eps_e} \ N_{\rm target}^m \ \epsilon (\eps_e) \ .
\end{equation}
In the above expression, $l$ and $m$ run all neutrino types and 
possible reactions, respectively, and 
$\eps_\nu$ is neutrino energy, 
$dF/d\eps_\nu$ differential number flux of neutrinos per unit neutrino energy, 
$d\sigma/d\eps_e$ differential
cross section, $N_{\rm target}$ total number of target particles
in the detector, and $\epsilon (\eps_e)$ the detection efficiency
(100 \% if $\eps_e > \eps_e^{th}$).
Although the sample data set generated from the MC code includes all
reactions in Eq. (\ref{eq:anue-p})--(\ref{eq:anue-O}), we analyze
the data considering only the dominant $\anue p$ reaction for simplicity.
In this approximation, the differential cross section is (Vogel 1984;
Burrows 1988)
\begin{equation}
 \frac{d\sigma_{\anue p}}{d\eps_e} = \frac{1}{4} \sigma_0 (1 + 3 \alpha^2)
 (1 + \delta_{\rm WM}) \ \frac{\eps_e p_e c}{(m_e c^2)^2} \
 \delta(\eps_\nu - \eps_e - \Delta_{np}) \ ,
\end{equation}
where $\sigma_0 = (2 G_F m_e \hbar)^2 / \pi c^2$, $\alpha$ the
axial-vector coupling constant ($\sim -1.26$), $p_e$ positron
momentum, and $\Delta_{np}$ the neutron-proton mass difference (1.294 MeV).
The weak-magnetism correction, $\delta_{\rm WM}$, is approximately
$-0.00325 \ (\eps_\nu - \Delta_{np}/2)$/MeV.
The $\anue$ number flux, $dF_{\anue}/d\eps_\nu$, is expressed as 
\begin{equation}
\frac{dF_{\anue}}{d\eps_\nu} = \frac{L_{\anue}}{4 \pi D^2
  T_{\anue}^4 F_3(\eta)} \ \frac{\eps_\nu^2}{e^{\beta (\eps_\nu -
    \mu)}+1} \ ,
\end{equation}
where $\mu$ is the chemical potential, $\beta = T_{\anue}^{-1}$,
$\eta = \mu / T_{\anue}$, (the Boltzman constant is set to the unity)
and $F_n (\eta)$ is defined as
\begin{equation}
F_n(\eta) \equiv \int_0^\infty \frac{x^n}{e^{x - \eta} + 1} dx \ .
\end{equation}
In the following analysis we set $\eta = 0$ and 
assume that the distance to the supernova is known.
If the distance is uncertain in practical analysis in the
future, the luminosity should be replaced by the bolometric flux,
i.e., $F_{\anue} = L_{\anue}/4 \pi D^2$.

\begin{figure}
  \begin{center}
    \leavevmode\psfig{figure=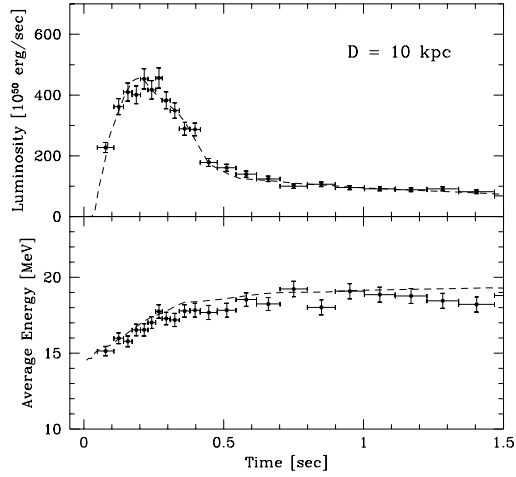,width=7cm}
  \end{center}
\caption{
Data points are
results of the time-binning analysis for a set of MC data of the SK detection
of a supernova at the Galactic center ($D$ = 10 kpc). The luminosity
and average energy are those of $\anue$'s above
the threshold energy, $\eps_e^{th} + \Delta_{np}$ =
11.3 MeV. The vertical error bars attached on the data points
indicate statistical 1 sigma errors,
while the horizontal bars represent the bin width.
The dashed lines are luminosity and average energy of the numerical 
supernova model (also above $\eps_e^{th} + \Delta_{np}$) from which the MC data
are generated. One time-bin includes about 400 events.}
\label{fig:d10-bin}
\end{figure}

\begin{figure}
  \begin{center}
    \leavevmode\psfig{figure=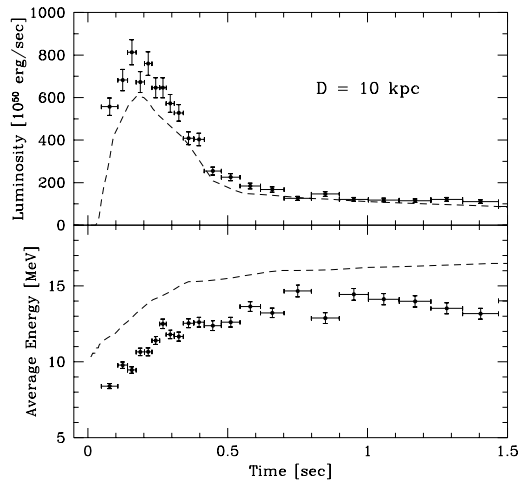,width=7cm}
  \end{center}
\caption{
The same as Fig. \protect\ref{fig:d10-bin}, but for
the luminosity and average energy of $\anue$'s in the whole energy
range including the lower energy range below $\eps_e^{th} + \Delta_{np}$
= 11.3 MeV.}
\label{fig:d10-bin-whole}
\end{figure}

We show a result of this time-binning
analysis on a sample data set ($D$ = 10 kpc) in
Fig. \ref{fig:d10-bin}.
Luminosity and average energy of $\anue$'s above 
$\eps_\nu > (\eps_e^{th} + \Delta_{np})$, which are
calculated from $L_{\anue}^i$ and $T_{\anue}^i$ obtained by
solving the Eqs. (\ref{eq:bin1}) and (\ref{eq:bin2}), are shown
as the data points.
The time bins are defined so that 
the event numbers in each bin are the same among the bins.
The number of events in one bin is $\sim$ 400 in this figure.
The statistical errors can approximately be estimated as
\begin{eqnarray}
{\rm Err}(L_{\anue}) &=& \frac{1}{\sqrt{N_i}} \qquad (\%) \ , \\
{\rm Err}(T_{\anue}) &=& \frac{1}{\sqrt{N_i}} \ 
 \frac{(\sigma_\eps)_i}{(\bar{\eps}_e)_i} \qquad (\%) \ ,
\end{eqnarray}
where $(\sigma_\eps)_i$ is the observed standard deviation of $\eps_e$ 
in the $i$-th bin. The error in luminosity is inferred from Poissonian
statistics and that in temperature from the statistical 1 $\sigma$ error
in the observed $(\bar{\eps}_e)_i$. Dashed lines are luminosity and
average energy of the numerical supernova model above $\eps_\nu
= (\eps_e^{th} + \Delta_{np}$).
It should be noted that we cannot know the offset time of the first
event in a future detection, but for the purpose of comparison,
we plot the fitted results in the same time coordinate with the
numerical supernova model unless otherwise stated.
The obtained luminosity and average energy well agree with those of
the original numerical supernova model. Although the fitted average
energy seems systematically lower than the original because of
the deviation from the FD distribution or contamination of reactions
other than $\anue p$, this systematic error is sufficiently small.
However, it should be
noted that significant systematic errors emerge when we 
extrapolate the obtained neutrino spectrum down to lower energy range below 
the threshold with the assumed FD distributions. 
Figure \ref{fig:d10-bin-whole} is the same as
Figure \ref{fig:d10-bin}, but for the whole energy range including
$\eps_\nu < (\eps_e + \Delta_{np})$. It is clear that an analysis
assuming the FD distribution leads to systematic overestimation
in luminosity and underestimation in average energy, because the
spectrum with obtained $T_{\anue}^i$ significantly overestimates
the neutrino flux below the threshold energy. This comes from the fact that
spectrum of numerical supernova models is `pinched', i.e.,
deficient in both low- and high-energy range compared to the pure FD 
distribution.

The features of the delayed mechanism discussed in \S 
\ref{section:simulation} can be seen clearly by this
simple analysis. Therefore a supernova at the Galactic center 
is near enough to get information for the explosion mechanism.
Now let us consider a case when a supernova is more
distant from the Earth. Figure \ref{fig:lmc-bin} is
the same as Figure \ref{fig:d10-bin}, but for a supernova
at the LMC ($D$ = 50 kpc). The event number in each time bin
is about 30. The statistical errors become larger and the features
of the explosion mechanism are obscured by the bin width and
statistical errors. It is difficult to reconstruct a smooth time
evolution from these discrete data points. Here we point out that 
although the time-binning analysis is simple and easy, this 
method loses some important information and 
we can reproduce a smooth evolution more efficiently by other methods
of analysis. In the next section, we discuss the information loss in the
time-binning analysis and propose a new method which uses a likelihood
function and cubic-interpolation.

\subsection{Likelihood Analysis with Cubic-Spline Interpolation}
The time-binning analysis is simple and clear, but it loses some
important information in the following two points. First,
the detection time of each event is smoothed out in the time bin
in which the event is included.
Generally water \v{C}erenkov detectors have fairly good time resolution
(much better than msec), but the time-binning analysis 
cannot extract any information on the time scale shorter than
the bin width. Second, the events in a bin are treated independently
of those in other bins, and this leads to statistical fluctuation
in the obtained results between neighboring bins. If we assume the time
evolution is smooth, we can suppress the statistical fluctuation
to some extent by smoothing. We consider here an analysis method
which is more effective in the above two points
than the time-binning analysis.

In order not to lose any information about detection time of
each event, we use the well-known maximum likelihood analysis,
where the likelihood function for the analysis of supernova neutrinos
is given as
\begin{equation}
  {\cal L} \equiv \log L = \sum_{j=1}^{N_{obs}} \ln \left( \frac{d^2N
  (t^j, \eps_e^j)}{dtd\eps_e} \right)  - \int_{t_l}^{t_u}dt 
  \int_{(\eps_e)_l}^{(\eps_e)_u}d\eps_e \frac{d^2N}{dtd\eps_e} \ ,
\end{equation}
where $t^j$ and $\eps_e^j$ are detection time and energy of $j$-th
event, $N_{\rm obs}$ the total number of events, and
$t_l$, $t_u$, $(\eps_e)_l$, and $(\eps_e)_u$ are the boundaries of 
the analysis region in the ($t$, $\eps_e$) space, which can be set arbitrarily.
[For the derivation of this function, see, e.g., Loredo \& Lamb (1989).]
Because we are interested in the time evolution, we assume the FD
distribution with $\eta = 0$ for energy spectrum. Then expected
$d^2N(t, \eps_e)/dtd\eps_e$ can be calculated if the time evolution
of $L_{\anue}$ and $T_{\anue}$ is given. Generally this time evolution
is modeled by some analytic functions, e.g., exponential or power-law
decay, but the time evolution with which we are concerned cannot
be modeled with any simple functions. We therefore set some grid
points in the time coordinate, $t_k$, ($k = 1,\ldots, N_{\rm grid}$)
and $L_k$ and $T_k$
are defined as $L_{\anue}(t_k)$ and $T_{\anue}(t_k)$, respectively.
We model $L_{\anue}(t)$ and $T_{\anue}(t)$ by the natural-cubic-spline
interpolation in order to reproduce a smooth evolution.
The merit of use of this interpolation is not only that
we can get a smooth time evolution, but the statistical fluctuation
is also suppressed. The reason is as follows.
Most methods for smooth interpolation, including the
natural-cubic-spline, are 
generally unstable to random fluctuation between the
neighboring grids, or to a random pattern.  If the values of
$L_k$ and $T_k$ become unrealistically random, values of $L_{\anue}$ 
and $T_{\anue}$
at points between the defined grids will become artificially
oscillatory and then the likelihood
function evaluated with cubic-spline interpolation will become lower.
Therefore this method is expected to suppress unnecessary statistical
fluctuation in time evolution.

\begin{figure}
  \begin{center}
    \leavevmode\psfig{figure=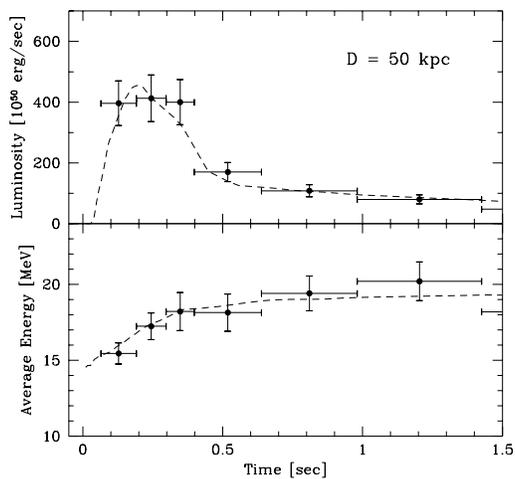,width=7cm}
  \end{center}
\caption{
The same as Fig. \protect\ref{fig:d10-bin}, but for a
supernova at the Large Magellanic Cloud ($D$ = 50 kpc). One time-bin includes
about 30 events.}
\label{fig:lmc-bin}
\end{figure}

\begin{figure}
  \begin{center}
    \leavevmode\psfig{figure=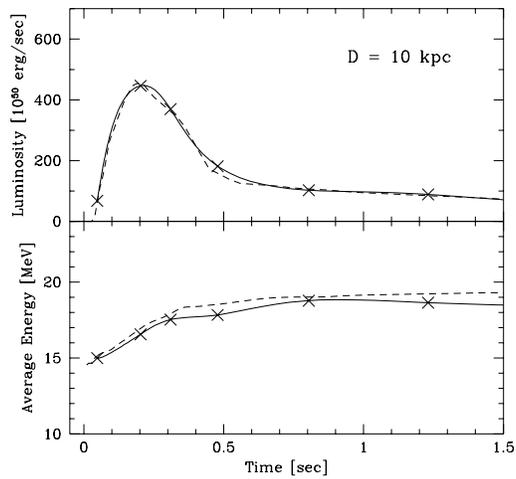,width=7cm}
  \end{center}
\caption{
The result of the likelihood analysis for a set of MC data
of the SK detection
of a supernova at the Galactic center ($D$ = 10 kpc). The luminosity
and average energy are those of $\anue$'s above
the threshold energy, $\eps_e^{th} + \Delta_{np}$ = 11.3 MeV.
The crosses are the grid points on the time coordinate, and the 
solid lines are the cubic-spline interpolation (see text). 
The dashed lines are luminosity and average energy of
the numerical supernova model (also above $\eps_e^{th} + \Delta_{np}$)
from which the MC data
are generated. The number of events per one grid point is about 800.}
\label{fig:d10-ncs}
\end{figure}

\begin{figure}
  \begin{center}
    \leavevmode\psfig{figure=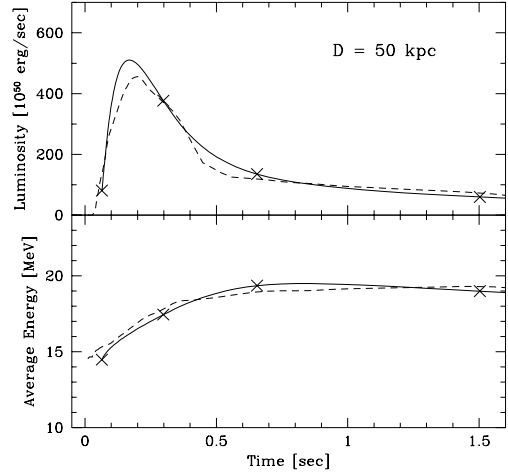,width=7cm}
  \end{center}
\caption{
The same as Fig. \protect\ref{fig:d10-ncs}, but for a
  supernova at the LMC ($D$ = 10 kpc). The number of events per one
  grid is about 60.}
\label{fig:lmc-ncs5}
\end{figure}

\begin{figure}
  \begin{center}
    \leavevmode\psfig{figure=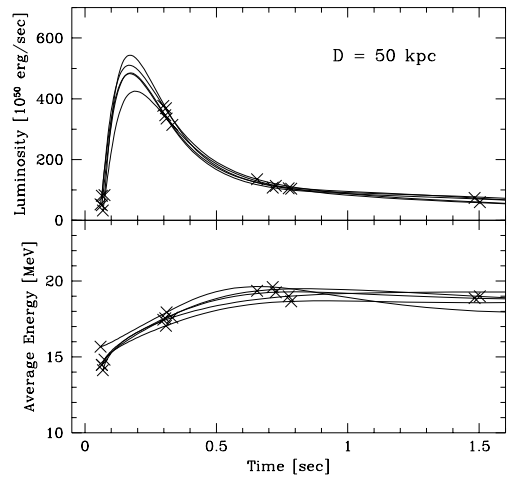,width=7cm}
  \end{center}
\caption{
The same as Fig. \protect\ref{fig:lmc-ncs5}, but for
other five sets of MC data generated from the same numerical supernova 
model. Statistical fluctuations in the obtained
results can be seen.}
\label{fig:lmc-ncs5-5mc}
\end{figure}

Now we can find the best-fit time evolution in $L_{\anue}$ 
and $T_{\anue}$ by searching maximum of ${\cal L}$ in the model
parameter space, $\{L_k, T_k \}$. We have adopted this analysis
method to the MC data for a supernova at $D$ = 10 and 50 kpc,
and the results about the luminosity and average energy above
the threshold energy ($\eps_e^{th} + \Delta_{np}$)
are shown in Figs. \ref{fig:d10-ncs} and
\ref{fig:lmc-ncs5}. We set the time grids so that 
the event numbers per one time grid is $\sim$ 800 and $\sim$ 60
for $D$ = 10 and 50 kpc, respectively, and these grids are expressed
as crosses in the figures.  The best-fit curves of
both cases (solid lines) well reproduce the smooth time evolution of 
the original numerical supernova model (dashed lines). These results should be
compared with those of the time-binning analysis shown in
Figs. \ref{fig:d10-bin} and \ref{fig:lmc-bin}, and it can be seen
that this method is useful to reconstruct a smooth and natural 
time evolution especially when a supernova is distant and available
events are fewer. For the $D=10$ kpc case
the difference of the two methods is 
not so important because of rich statistics, but the initial
steep rise of the luminosity curve can better be reproduced by 
the likelihood analysis than the time-binning analysis.
In order to show the statistical uncertainties in the above analysis,
we make other five data sets of MC simulation from the same numerical
supernova model for the $D$ = 50 kpc
case, and do the same analysis for all of them. The results are shown
in Fig. \ref{fig:lmc-ncs5-5mc}, and this figure should be compared
to the statistical errors in Fig. \ref{fig:lmc-bin}. 

\begin{figure}
  \begin{center}
    \leavevmode\psfig{figure=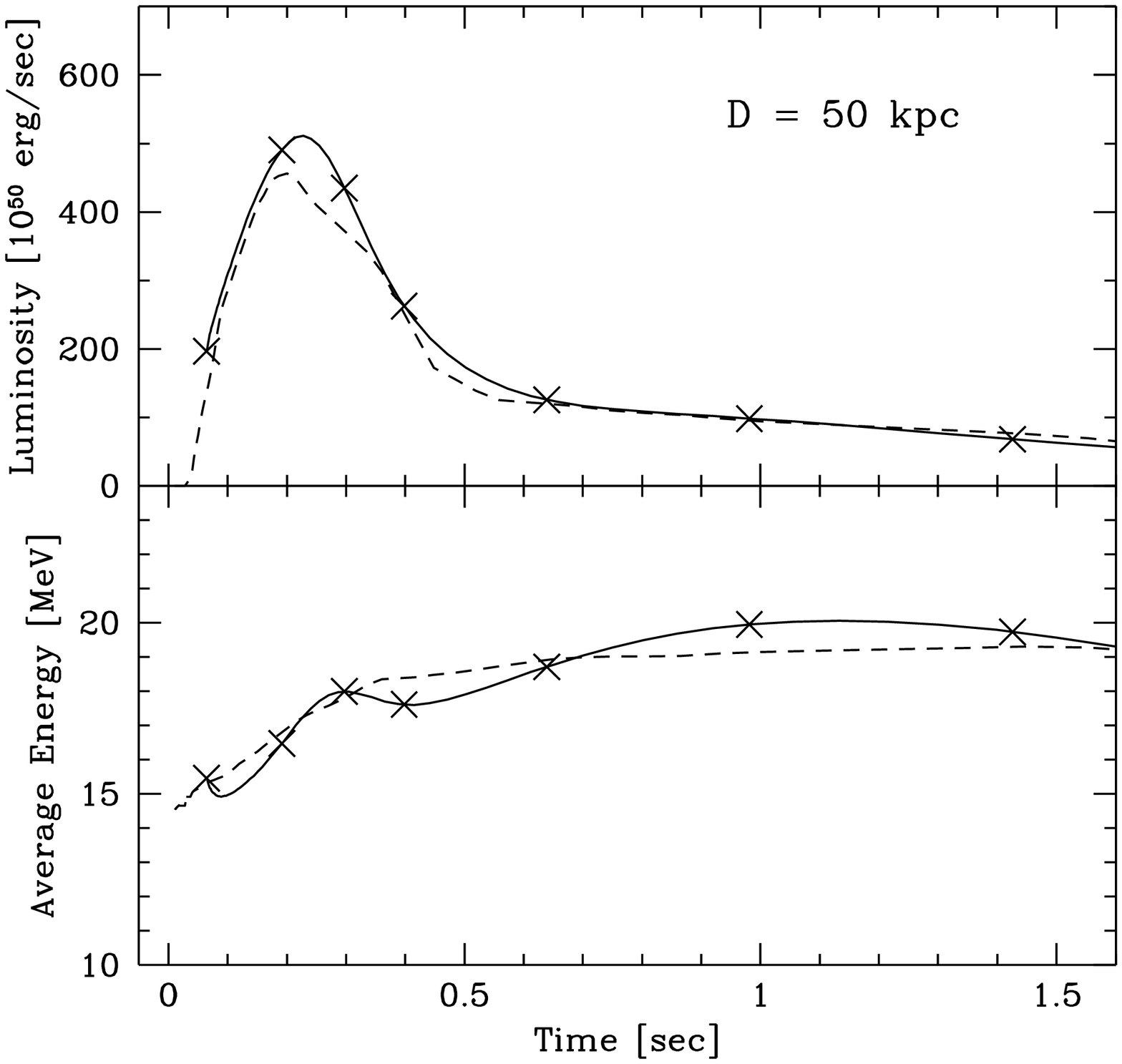,width=7cm}
  \end{center}
\caption{
The same as Fig. \protect\ref{fig:lmc-ncs5}, but the number
  of grid points is increased by a factor of 2.}
\label{fig:lmc-ncs10}
\end{figure}

\begin{figure}
  \begin{center}
    \leavevmode\psfig{figure=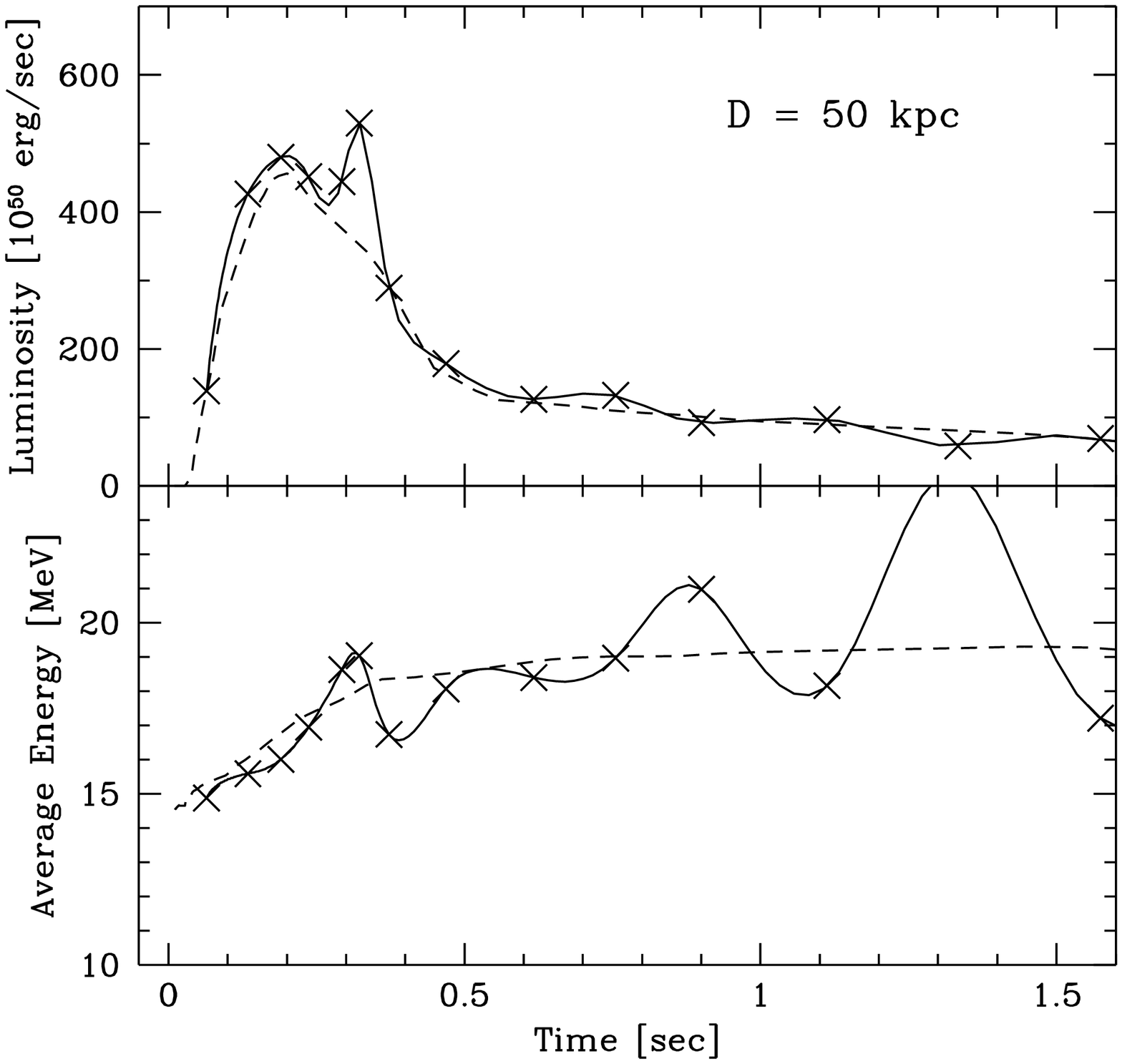,width=7cm}
  \end{center}
\caption{
The same as Fig. \protect\ref{fig:lmc-ncs5}, but the number
  of grid points is increased by a factor of 4.}
\label{fig:lmc-ncs20}
\end{figure}

The number of grid points defined on the time coordinate
is crucially important in the above analysis.
We should choose a sufficient number of grids which can resolve
the time scale in which we are interested, but
it is apparent that if we set too many time grids, the statistical
fluctuation per one grid will become larger and obtained results become
unstable. In Figs. \ref{fig:lmc-ncs10} and \ref{fig:lmc-ncs20},
we show the result of the same analysis about the $D = 50$ kpc data
with different number of grids; the number of grid points is increased
by factors of 2 and 4 from that in Fig. \ref{fig:lmc-ncs5},
respectively. The instability can be seen clearly, and
therefore discrimination of the true time evolution from
sham evolution due to statistical fluctuation is important.
In a practical analysis in the future,
we should try various intervals of grids taking account the 
statistical errors estimated by the time-binning analysis.
The goodness-of-fit test, which we discuss in the next section,
will also help us to discriminate a true evolution from statistical
fluctuations. If we get sufficient goodness-of-fit with a given 
number of grids, the analysis with more grids is unnecessary.

\subsection{Goodness-of-Fit test and Deviation from the FD distribution}
Generally a statistical analysis includes the following three
procedures: 1) find the best fit parameters, 2) estimate the
statistical errors, and 3) check whether the best-fit model is consistent
with the observed data. Here we consider the 
third procedure, a so-called goodness-of-fit (GOF) test.
If a likelihood function is given with some assumptions about the
model, it is rather a straightforward process to find the best-fit
parameters. However, the likelihood analysis itself does not verify
the assumed model, and therefore we have to check whether
the observed data naturally come out from the assumed model with best-fit
parameters. We use the two-dimensional version of the 
Kolmogorov-Smirnov (KS) test (Peacock 1983; Fasano \& Fianceschini 1987)
as a tool of the GOF test. The KS measure $D_{\rm KS}$, which is a measure of 
deviation of the observed data from the best-fit model, is defined as follows.
First, we define the
expected and observed fraction of events in the four quadrants of 
$(t, \eps_e)$ space: 
\begin{equation}
f^l_{\rm exp}(t, \eps_e) = \frac{1}{N_{\rm exp}}
  \int\!\!\!\int_l dt'd\eps_e' \frac{d^2N(t', \eps_e')}{dtd\eps_e} \ ,
\end{equation}
and $f^l_{\rm obs}(t, \eps_e) = N_{\rm obs}^l(t, \eps_e)/N_{\rm obs}$,
where $l$ (= 1, 2, 3, and 4) denotes the four quadrants: 
$(t'< t, \eps_e' < \eps_e)$, $(t'> t, \eps_e' < \eps_e)$, 
$(t'< t, \eps_e' > \eps_e)$, and $(t'> t, \eps_e' > \eps_e)$.
The quantity $N_{\rm obs}^l (t, \eps_e)$ is the number of events in the $l$-th
quadrant, and $N_{\rm exp}$ is the total expected number of events
in the whole analysis region. 
Then $D_{KS}$ is defined as the maximum of the absolute difference of $f^l_{\rm
exp}$ and $f^l_{\rm obs}$, i.e.,
\begin{equation}
D_{KS} = \max_{l, \ t, \ \eps_e} |f^l_{\rm exp}(t, \eps_e) - 
 f^l_{\rm obs}(t, \eps_e)| \ .
\end{equation}
We can check the consistency between the observed data and the
best-fit model by comparing $D_{KS}$ of the observed data and
probability distribution of $D_{KS}$ expected from the best-fit
model. The probability distribution of $D_{KS}$ 
is unknown in this two dimensional case, and
we have to estimate this by a number of MC simulations.

\begin{figure}
  \begin{center}
    \leavevmode\psfig{figure=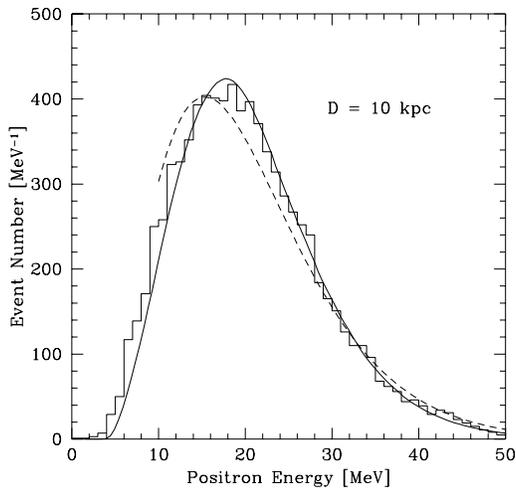,width=7cm}
  \end{center}
\caption{
Time-integrated 
energy distribution of events at the SK for a supernova
at the Galactic center ($D$ = 10 kpc). The histogram is the MC data
analyzed, and the solid line is the distribution of $\anue p$ events
expected from the numerical supernova model from which the MC data
are generated. The dashed line is the best-fit distribution 
determined by the likelihood analysis (Fig. \protect\ref{fig:d10-ncs})
assuming the Fermi-Dirac (FD) distribution ($\mu = 0$). The difference of
the FD fit and the MC data is clear.}
\label{fig:dNde-d10}
\end{figure}

\begin{figure}
  \begin{center}
    \leavevmode\psfig{figure=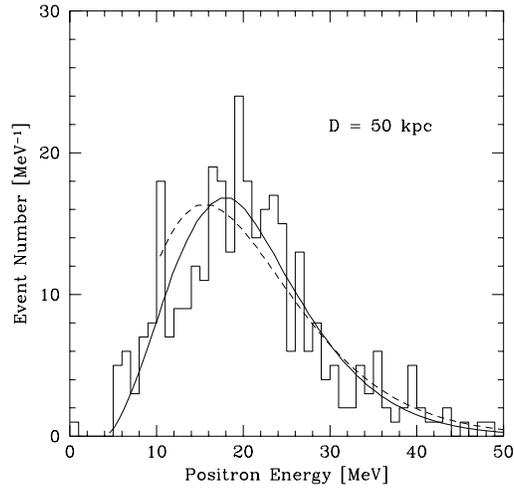,width=7cm}
  \end{center}
\caption{
 The same as Fig. \protect\ref{fig:dNde-d10}, but for
  a supernova at the LMC ($D$ = 50 kpc). The deviation of
  the FD fit from the MC data cannot be distinguished from statistical
  fluctuations.}
\label{fig:dNde-d50}
\end{figure}

We have applied this test on the results obtained by the likelihood 
analysis in the previous section for both the $D$ = 10 and 50 kpc 
cases (Fig. \ref{fig:d10-ncs} and \ref{fig:lmc-ncs5}), by using the 
probability distribution of $D_{KS}$ obtained by 100 MC simulations. 
It is found that the best-fit model and the MC data are statistically
consistent for
the $D$ = 50 kpc case, but inconsistent for the $D$ = 10 kpc case
with more than 99 \% C.L.  Because the time evolution is considered
to be modeled appropriately, this suggests that the inconsistency
comes from the assumption of the FD distribution in energy spectrum.
This means that, in other words, we can distinguish the difference of
the real spectral shape of supernova $\anue$'s 
from pure FD distributions when a supernova occurs nearer than 10 kpc.
In order to demonstrate this, we plot the time-integrated energy
spectrum of events expected from the best-fit model obtained by the
likelihood analysis assuming the FD distribution, 
by the dashed line in Fig. \ref{fig:dNde-d10} for the $D$ = 10 kpc
case. The histogram is the analyzed MC data and the solid line is the 
expected spectrum of the numerical model from which the MC data are 
generated. The difference of the dashed line from the histogram
is clearly discernible; overestimation at $\eps_e < 18$
MeV and underestimation at $\eps_e > 18$ MeV. This deviation should be 
considered as a prediction of a standard picture of neutrino emission
from collapse-driven supernovae and can be tested in a future observation.
Figure \ref{fig:dNde-d50} is the same as Figure \ref{fig:dNde-d10},
but for the $D$ = 50 kpc case. In this case it seems difficult to see
the deviation from the FD distribution, and this is consistent
with the results of the GOF test.

\section{Discussion}
\label{section:discussion}
Neutrino-driven Rayleigh-Taylor instabilities between the stalled
shock wave and the neutrinospheres are generic feature of
collapse-driven supernovae (Bethe 1990; Herant, Benz, \& Colgate 1992;
Herant et al. 1994; Burrows, Hayes, \& Fryxell 1995; Janka \&
M\"{u}ller 1996; Mezzacappa et al. 1996). Such instability ineviably
leads to convective motion in the energy gain region and asphericity
in the dynamics, that should have significant effects on the
explosion mechanism (see, e.g., Burrows 1997 for a review). 
Although the presented 1-D calculation takes 
account of convective motion by the mixing length theory, there
may be some effects that can not be covered by 1-D calculation.
It is interesting to consider the possible signatures of such 
asymmetry imprinted in neutrino emission, but unfortunately at
the present stage the effect of convective motion on the emergent
neutrino luminosity or spectrum is poorly known. 
Furthermore, rotation might tend to wash out otherwise detectable 
flux variations due to convective motion. In fact, the Crab pulsar
rotates at approximately 200 radians/sec, and this rotation will 
presumably smear out other underlying observable variations.

There are some interesting hints for finite neutrino masses, such as
the solar neutrino problem or the atmospheric neutrino anomaly.
Neutrino oscillations due to these possible masses might significantly
change the emergent spectrum of neutrinos. If the vacuum mixing angle
among the three generation of neutrinos is order unity, this leads
to the vacuum neutrino oscillation. Because supernova neutrinos come
out through very dense matter, it is also possible that the MSW neutrino
oscillation occurs such as $\nu_e \leftrightarrow \nu_\mu$
(Fuller et al. 1987). Under the direct 
mass hierarchy of neutrinos (i.e., $m_{\nu_e} < m_{\nu_\mu} < m_{\nu_\tau}$),
the MSW matter oscillation is relevant only for neutrinos and not for 
antineutrinos, but it is also possible that $\anue$'s, which we mainly
discussed in this work, experience resonant matter oscillation with
$\nu_\mu$ or $\nu_\tau$, due to
flavor-changing magnetic moment of Majorana neutrinos [spin-flavor precession,
see e.g., Totani \& Sato (1996)]. These phenomena might
significantly change the neutrino spectrum, and detectability of these
signature will be interesting topics in future work.

\section{Summary and Conclusions}
\label{section:summary}
We performed a statistical analysis for the future detection
of a supernova neutrino burst at $D$ = 10 kpc (the Galactic center)
and $D$ = 50 kpc (LMC) by the Super-Kamiokande detector,
by using a numerical supernova model and realistic Monte-Carlo (MC)
simulations of detection. We mainly discussed the detectability
of the signatures of the delayed explosion mechanism in the time evolution
of the $\anue$ luminosity and spectrum: a hump during the first
$\ltilde$ 0.5 second and following abrupt drop in the $\anue$
luminosity curve, and also corresponding spectral hardening. 
[It should be noted that
these signatures generally depend on the mass and internal structure
of the progenitor star.
The model used is for SN1987A, i.e., its progenitor is a
$\sim$ 20 $M_\odot$ main-sequence star.]  
The $\nu_e$ neutronization burst is considered to be more 
energetic for the delayed explosion and hence could be another clue
to the explosion mechanism. However our simulation of the delayed explosion
produces only about 5 scattering events due to neutronization burst
when $D$ = 10 kpc, and it seems difficult to distinguish 
clearly the burst from $\anue p$ events.

We analyzed MC data
generated from the numerical supernova model in the energy range
above 10 MeV (to avoid background noise), and found the following
results. 1) The signatures of the delayed explosion in $\anue$
luminosity curve and spectral evolution are clearly
discernible for the $D$ = 10 kpc case, and moreover, the difference of
the real energy spectrum from pure Fermi-Dirac (FD) distribution 
can also be observable.
2) For the $D$ = 50 kpc case, the signature of the delayed explosion
is still observable, but statistical fluctuation is too large to
distinguish the deviation from the FD distribution. 
These results suggest that we will be able to distinguish the two
proposed explosion mechanisms if a supernova occurs in Our Galaxy or
Magellanic Cloulds in the near future. The deviation
from the FD distribution would, if observed, provide an important test
for the standard picture of neutrino emission from collapse-driven
supernovae. The FD fitting leads to significant overestimation of flux
in lower energy region ($\ltilde$ 15 MeV), and this results in 
overestimation in luminosity and underestimation in average energy.
We should be careful for this in a future analysis
when the obtained results are extrapolated down to
the lower energy region below the threshold.

Time-binning analysis is the simplest method to reconstruct the
time evolution of $\anue$ flux and spectrum, but this method loses
some important information and hence is not maximally effective. Therefore
we proposed a method for reconstruction of the time evolution, 
which gives a smoother time evolution
and smaller statistical errors than the simple time-binning analysis.
This method is based on the likelihood analysis, and its
characteristic is the use of cubic-spline interpolation to express
the time evolution of $\anue$ luminosity and effective temperature with
some selected grid points on the time coordinate. 
The likelihood analysis does not lose any detection time
information and the stability of the cubic-spline interpolation
to random fluctuation suppresses statistical fluctuations. This method is 
useful especially when available number of events is relatively small,
e.g., a supernova in the LMC or SMC.

This work has been supported in part by the Grant-in-Aid (KS) for the 
Center-of-Excellence Research No. 07CE2002 and (TT) for the Scientific
Research Fund No. 3730 of the Ministry of Education, Science, and
Culture in Japan.
This work was also supported by (HED and JRW) DOE Contract No.
W-7405-ENG-48 
and (JRW) NSF Grant No. PHY-9401636.

\newpage

\end{document}